\documentclass[%
twocolumn,
superscriptaddress,
amsmath,amssymb,
aps,
prd,
floatfix,
]{revtex4-2}

\usepackage{graphicx}
\usepackage{dcolumn}
\usepackage{bm}
\usepackage{lineno,siunitx,placeins,subfigure,float,amsmath}
\usepackage[utf8]{inputenc}  
\usepackage[T1]{fontenc}     
\usepackage{lmodern}

\begin{document}
\preprint{APS/123-QED}
\title{Strain sensitivity enhancement in a Grover-Michelson interferometer}

\author{Anthony D. Manni}
 \email{admanni@bu.edu}
 \affiliation{Dept. of Electrical and Computer Engineering, Photonics Center, Boston University, 8 Saint Mary's Street, Boston, MA 02215, USA}
\author{Christopher R. Schwarze}
 \email{crs2@bu.edu}
 \affiliation{Dept. of Electrical and Computer Engineering, Photonics Center, Boston University, 8 Saint Mary's Street, Boston, MA 02215, USA}

\author{David S. Simon}
 \email{simond@bu.edu}
 \affiliation{Dept. of Physics and Astronomy, Stonehill College, 320 Washington Street, Easton, MA 02357}
 \affiliation{Dept. of Electrical and Computer Engineering, Photonics Center, Boston University, 8 Saint Mary's Street, Boston, MA 02215, USA}

\author{Abdoulaye Ndao}
 \email{a1ndao@ucsd.edu}
 \affiliation{Dept. of Electrical and Computer Engineering, University of California San Diego, Engineers Lane, San Diego, CA 92161, USA}

\author{Alexander V. Sergienko}
 \email{alexserg@bu.edu}
\affiliation{Dept. of Physics, Boston University, 590 Commonwealth Ave., Boston, MA 02215, USA}
\affiliation{Dept. of Electrical and Computer Engineering, Photonics Center, Boston University, 8 Saint Mary's Street, Boston, MA 02215, USA}

\begin{abstract} 
The Michelson interferometric phase detection resolution can be enhanced by replacing conventional beam splitters with novel directionally unbiased four-port scatterers, such as Grover coins. We present a quantitative analysis of the noise-to-signal ratio of sideband frequencies generated by gravitational wave-induced phase perturbations in a Grover-Michelson interferometer (GMI). We discuss the principles of GMI signal enhancement and demonstrate how combining this configuration with additional light-recycling arrangements further enhances the performance.

\end{abstract}

\maketitle

\section{Introduction}

Long-baseline, terrestrial laser interferometers such as Advanced LIGO (aLIGO) in the U.S. and VIRGO in Europe rely on minute displacements of suspended mirrors to detect small perturbations to space-time due to passing gravitational waves (GWs) \cite{aligo,avirgo,LIGO:GWobs,GWdet:review}. The noise-to-signal ratio (NSR) of the frequency sidebands generated by such waves depends on the power and storage time of the light inside the device, dictating the smallest detectable GW amplitude \cite{michelson:gw}. State-of-the-art detectors employ dual-recycled Fabry-Perot Michelson interferometers \cite{Meers:recycling,Fritschel:powerrec,Gray:recycling,Strain:dualreccontrol}, where power and signal recycling mirrors are placed at the input and output of the central beam splitter, respectively, and Fabry-Perot cavities are placed in each arm of the Michelson interferometer (MI) to increase the effective path length and recirculating power.
The broadband gain in sensitivity of a dual-recycled interferometer is fixed by the finesse and loss of light recycling cavities, while the amplitudes of the high-frequency sidebands are limited by the storage time and coherence length of the carrier wave \cite{dualrec:susp,Strain:dualrecexp}. 
In addition to quantum shot noise, major contributions to signal noise include the radiation pressure on the suspended end mirrors due to the many kilowatts of recirculating power in the cavities, frequency and intensity noise of the laser source, and coating thermal noise (CTN) \cite{Khalili:rpnoise,Harry:tfthnoise}. These noise sources limit the sensitivity of terrestrial interferometers to low-frequency GWs generated by compact binary systems, supernovae, and pulsars \cite{GWdet:review}.
Recent work by Krenn, Drori, and Adhikari involved the digital discovery of GW detector topologies with enhanced sensitivity compared to the state-of-the-art by using artificial intelligence and numerical optimization \cite{krenn:UIFO}. The authors produced fifty successful designs that tended to be resource-heavy and governed by physics that is less intuitive than in first-principles methods.
\\Another approach to designing new GW detectors involves the use of novel interferometers based on directionally unbiased multiports, which support tunable sensitivity by adjusting the positions of the two end mirrors \cite{schwarze_ub-sens-thry,Schwarze:gsi}. In such devices, the central beam splitter of, e.g., a Michelson or Sagnac interferometer, is replaced by a four-port Grover coin, which enables the coupling between mirror arms by reflecting light back through entry ports \cite{shuto_entropy,shuto_clusterwalk,shuto_hamil,shuto_qrouter}. A recent benchtop demonstration of a Grover-Michelson interferometer (GMI) showed an increase in output sensitivity to differential arm length of more than an order of magnitude compared to a conventional MI \cite{Schwarze:gmi_exp}. 
\\As will be seen below, the optical power recirculating in a GMI is proportional to the common and differential arm lengths and can be increased by several orders of magnitude over the input power. In Section II, we discuss the operating principle of a Grover-Michelson interferometer and derive the device's cavity field equations. In Section III, we present expressions for the amplitude and NSR of sideband frequencies generated by GWs within a GMI and compare them with those of a conventional MI. In Section IV, we implement the \textsc{finesse} Python package to numerically analyze the NSR of a GMI in the presence of radiation pressure, and compare it with that of a simplified layout of aLIGO \cite{finesseref}. In Section V, we study how the performance of the GMI can be enhanced with additional light recycling configurations, and show that both power and signal recycling, separately, can achieve reductions in NSR compared to a standard GMI at frequencies below 25 Hz. In Section VI, we discuss the practical aspects of implementing a long-baseline GMI-based gravitational wave detector (GWD) and provide operational targets for the laser frequency and intensity noise that will enable the GMI's enhancement in strain sensitivity.

\section{Grover-Michelson Interferometer}\label{sec:ii}
\subsection{Operating Principle}
\begin{figure}[ht!]
\includegraphics[width=.8\linewidth]{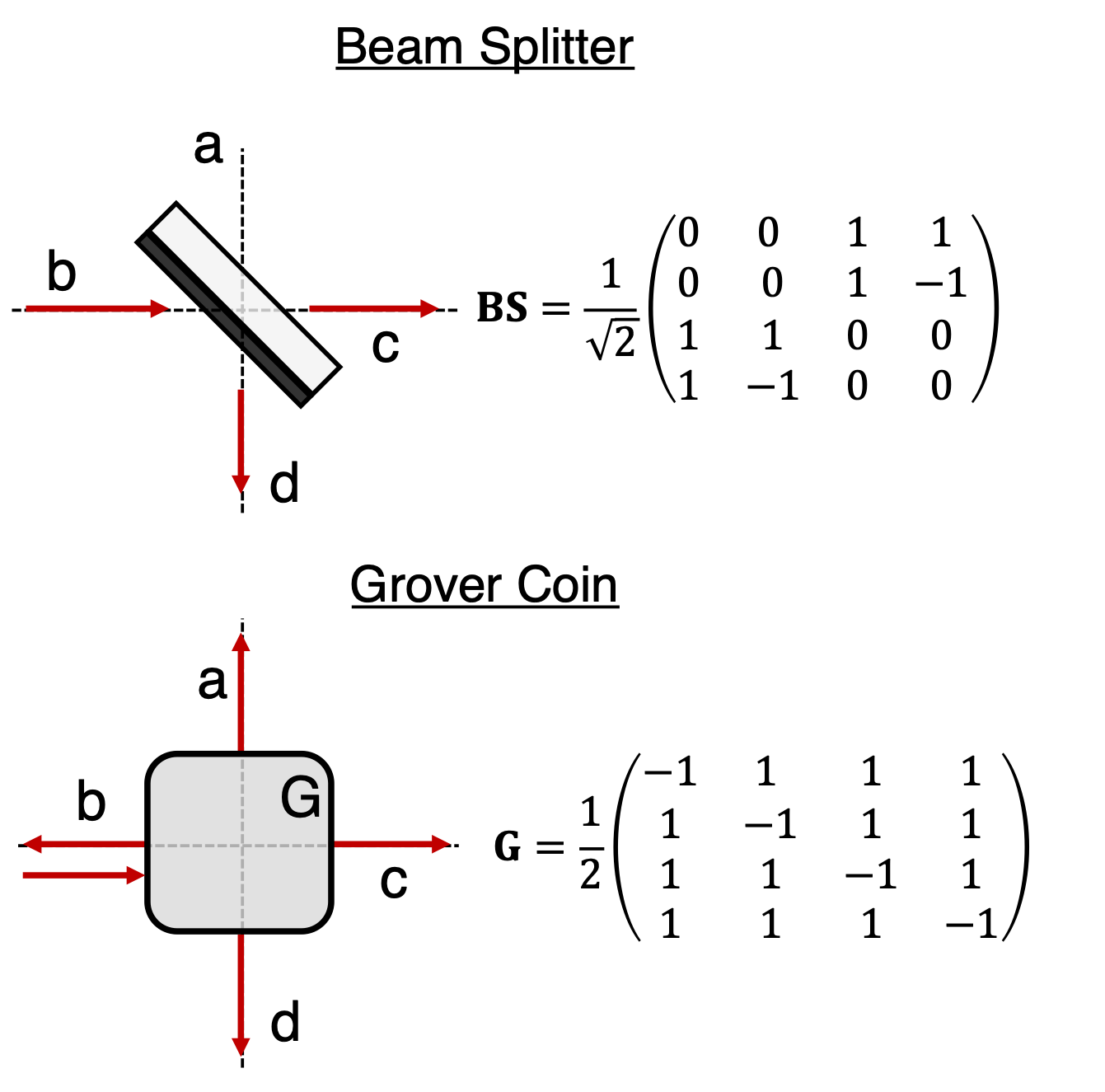}
\caption{\label{fig:gcbs} Graphical representation of the actions of a dielectric beam splitter and a Grover coin. Their corresponding 4x4 unitary matrices are provided next to each graphic. The dielectric beam splitter routes incoming photons from one input to two output ports, with a relative phase determined by the orientation of the dielectric film. Due to Fresnel reflectance, the output amplitudes acquire a $\pi$-phase shift when incident on the film from air and no phase shift when incident through the bulk of the substrate. The Grover coin routes photons to all four output ports, regardless of the input port. Output amplitudes acquire a $\pi$-phase shift when reflected back to their input port.}
\end{figure}

\begin{figure}[b]
\includegraphics[width=\linewidth]{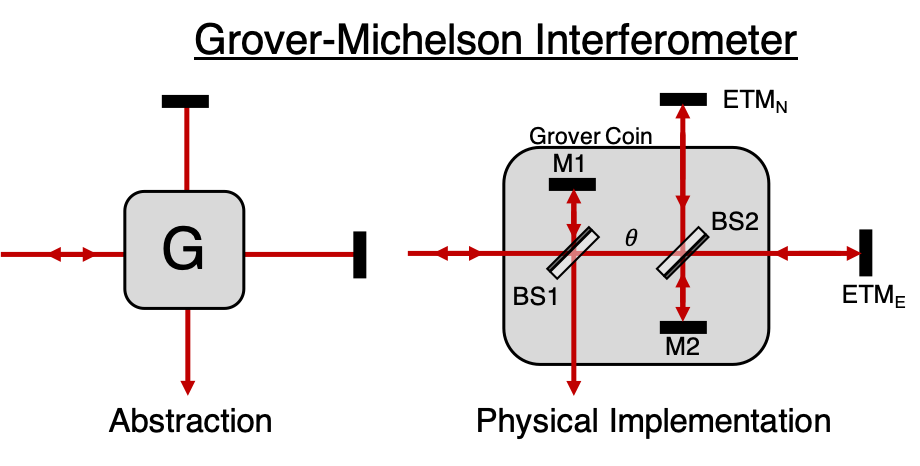}
\caption{\label{fig:graphic}Abstract representation (left) and physical implementation (right) of a Grover-Michelson interferometer for gravitational wave detection. A Grover coin can be decomposed into a combination of balanced beam splitters and mirrors with relative phase shifts all set to 0 mod 2$\pi$. The end test masses (ETMs) in the interferometer arms are suspended, weighing 40kg each.}
\end{figure}

The central component in a Michelson interferometer is the beam splitter, which can be thought of as a feedforward or ``directionally biased'' four-port, since a photon entering one port can only exit two of the four available ports. In contrast, a Grover coin provides input photons with equal probabilities of exiting all four of the ports, including the one through which it entered \cite{schwarze_ub-sens-thry}. Thus, the Grover-Michelson interferometer is constructed by replacing the central beam splitter of a traditional Michelson interferometer with a Grover coin \cite{Schwarze:gmi_exp}. Figure \ref{fig:gcbs} provides a graphical representation of the distinction between scattering properties of the two components. The unitary matrix of the beam splitter $BS$ is represented in the 4-mode basis to highlight the feedforward behavior in contrast to the directionally unbiased Grover coin.
\\The Grover coin scattering behavior is described by the unitary operator $G$,
\begin{equation}
    G = \frac{1}{2}\begin{pmatrix}
        -1 & 1 & 1 & 1 \\
        1 & -1 & 1 & 1 \\
        1 & 1 & -1 & 1 \\
        1 & 1 & 1 & -1
    \end{pmatrix} \;,
\end{equation}
where the minus signs along the diagonal result from reflected photon amplitudes acquiring a $\pi$ phase shift, i.e. reversing their direction of propagation. Labeling the ports a, b, c, d, then the Grover coin transforms an incoming photon in, e.g., port $a$ as follows:
\begin{equation}
    G\hat{a}^\dagger|0\rangle = \frac{1}{2}(-\hat{a}^\dagger + \hat{b}^\dagger + \hat{c}^\dagger + \hat{d}^\dagger)|0\rangle \;,
\end{equation}
where $\hat{a}^\dagger$ is the creation operator for a single photon in mode $a$, and so on. Placing mirrors at two of the output ports of the Grover coin results in a Grover-Michelson interferometer, as shown in Figure \ref{fig:graphic}, where the reflections in the Grover coin establish phase-dependent coupled cavities. 

 Schwarze et al. demonstrated that a Grover coin can be implemented in free space using a combination of two balanced beam splitters and two mirrors \cite{Schwarze:yc}. This embodiment is favorable in that it does not form any resonant cavities within the coin and can be made broadband and insensitive to polarization, depending on the beam splitters used. Throughout this paper, we consider plane monochromatic waves of a single polarization propagating through a Grover-Michelson interferometer, where the Grover coin is built with balanced, lossless, dielectric beam splitters (BSs) oriented such that the coated surfaces of BS1 and BS2 face the internal mirrors M1 and M2, respectively (see Fig. \ref{fig:graphic}). The northern and eastern ports of BS2 are connected to mirror-coated end test masses (ETMs), which are mechanically suspended to sample the effect of proper length distortion caused by passing GWs. Defining the two cavities as north ($N$) and east ($E$), the cavity relative field amplitude equations are (see Appendix \ref{app:cavs} for derivation)
\begin{equation}\label{eq:cavs}
    \frac{E_{N(E)}}{E_0} = \frac{e^{j\theta}\left(1-r_{E(N)}r_2e^{j(\phi_{E(N)}+\phi_2)}\right)}{2-r_2e^{j\phi_2}(r_Ne^{j\phi_N}+r_Ee^{j\phi_E})} \;,
\end{equation}
where $E_0$ is the incident carrier wave amplitude of wavelength $\lambda_0$ and wavenumber $k_0=2\pi/\lambda_0$, $j=\sqrt{-1}$ is the imaginary unit, $r_N(r_E)$ and $\phi_{N}=2k_0L_{N}(\phi_{E}=2k_0L_{E})$ are the amplitude reflectance of the north(east) end mirrors and roundtrip phases accumulated in the north(east) arms of lengths $L_N(L_E)$, respectively, $r_2$ is the reflectivity of the second internal Grover coin mirror, and $\phi_2=2k_0L_2$ is the roundtrip phase accumulated in the internal mirror M2 path.
\\
The normalized field transmitted through the bottom output port of the Grover coin, $\tilde{E}_t=E_t/E_0$, is (see Appendix \ref{app:b}),
\begin{equation}\label{eq:et}
\begin{aligned}
& \tilde{E}_t = \frac{e^{j2\theta}}{2}\left[\frac{r_Ne^{j\phi_N}+r_Ee^{j\phi_E}-2r_2r_Nr_Ee^{j(\phi_2+\phi_N+\phi_E)}}{r_2e^{j\phi_2}(r_Ne^{j\phi_N}+r_Ee^{j\phi_E})-2}\right] \\
& + \frac{1}{2}r_1e^{j\phi_1}
\end{aligned}
\end{equation}
where $r_1$ is the amplitude reflectance of internal mirror M1 and $\phi_1=2k_0L_1$ is the roundtrip phase accumulated in the M1 arm. In the case of a pure Grover coin, that is, a single scatterer represented by (1), the values $r_1e^{j\phi_1},\,r_2e^{j\phi_2},$ and $e^{j\theta}$ are set to 1. In practice, operation of a GMI in this configuration requires control of five different phases, three internal to the coin, and two external. Assuming unit reflectivity of the end mirrors, the transmitted field amplitude becomes identical to that derived in \cite{Schwarze:gmi_exp} using the quantum optical formalism, 
\begin{equation}
    \tilde{E}_t = \frac{1}{2}\left[\frac{e^{j\phi_N}+e^{j\phi_E}-2e^{j(\phi_N+\phi_E)}}{e^{j\phi_N}+e^{j\phi_E}-2}+1\right]\;,
\end{equation}
and similarly, the reflected field amplitude is
\begin{equation}
    \tilde{E}_r = \frac{1}{2}\left[\frac{e^{j\phi_N}+e^{j\phi_E}-2e^{j(\phi_N+\phi_E)}}{e^{j\phi_N}+e^{j\phi_E}-2}-1\right]\;.
\end{equation}
For input through the bottom port, the expressions for $\tilde{E}_t$ and $\tilde{E}_r$ are swapped. Note that the complex term in (5) is defined on the unit circle, and can thus be represented by a single phasor
\begin{equation}\label{eq:gammadef}
    e^{j\gamma(\phi_N,\phi_E)} \equiv \frac{e^{j\phi_N}+e^{j\phi_E}-2e^{j(\phi_N+\phi_E)}}{e^{j\phi_N}+e^{j\phi_E}-2}
\end{equation}
with the mapped nonlinear phase $\gamma(\phi_N,\phi_E)$ defined as $\arg\{e^{j\gamma}\}$ (see Appendix \ref{app:c}),
\begin{equation}\label{eq:gamma}
\begin{aligned}
    &\gamma \equiv\text{atan2}\big(\sin\Theta-\sin\phi_N-\sin\phi_E,
    \\&\;\cos\Theta-\cos\phi_N-\cos\phi_E + \frac{1}{2}[1+\cos\phi]\big)
\end{aligned}
\end{equation}
where $\Theta=\phi_N+\phi_E$, $\phi=\phi_N-\phi_E$, and atan2 is the two-argument arctangent. The transmitted and reflected electric fields may be redefined in terms of $\gamma$ as
\begin{equation}\label{etgamma}
    \frac{E_t}{E_0} = e^{j\gamma/2}\cos\left(\frac{\gamma}{2}\right)\;,
    \qquad \frac{E_r}{E_0} = je^{j\gamma/2}\sin\left(\frac{\gamma}{2}\right)\;.
\end{equation}
\\We see from \eqref{etgamma} that the interferometer is placed on a bright fringe, or maximum transmission, whenever $\gamma$ takes on a value of 0 mod $2\pi$, i.e., when the numerator in \eqref{eq:gamma} vanishes. This condition is met when the sum of the round-trip phases in the northern and eastern arms is equal to $2\pi n$, $n\in \mathbb{Z}$ (see Appendix \ref{app:tmax}). Figure \ref{fig:gamma} shows the function $\gamma(\phi_N,\phi_E)$ versus $\phi_E$ for different values of $\phi_N$, along with the transmitted intensity in the bottom port of the first beam splitter (BS1). The characteristic behavior of the Grover-Michelson interferometer is shown, where one mirror arm sets the shape of the response curve, while the other arm scans along it. This can also be expressed in relative terms by the smaller of the two differences in phases of maxima and minima. We identify the minima with $2\pi n$, but this is arbitrary fixing of a global phase. Repeated scanning of one phase allows determination of the other phase by locating the transmission peak and width. This differs from the behavior of the standard Michelson interferometer, where the second phase simply translates the curve without changing its shape \cite{Schwarze:gmi_exp}.
\begin{figure}[ht!]
\centering
\includegraphics[width=\linewidth]{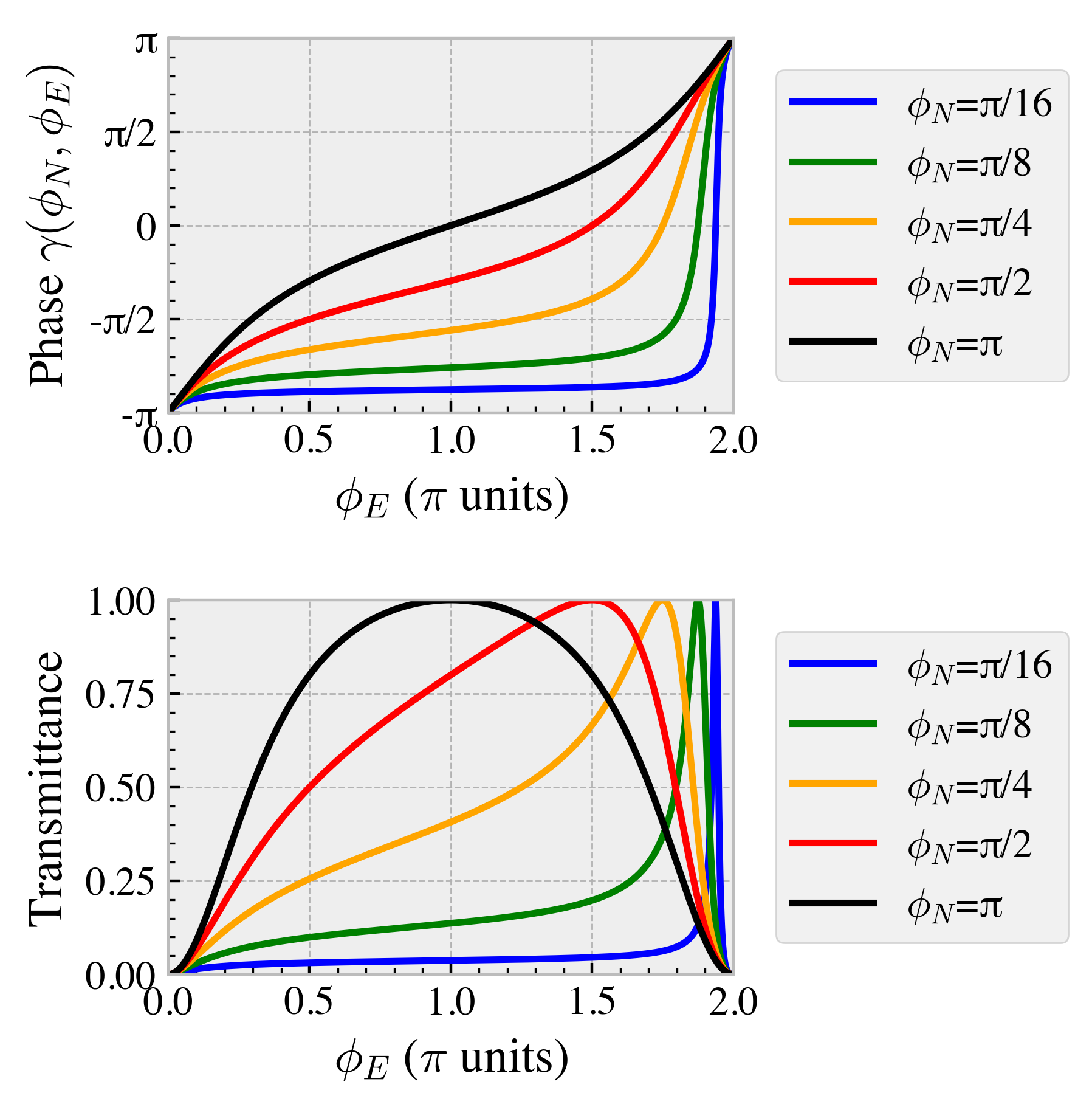}
\caption{\label{fig:gamma} Nonlinear phase function $\gamma(\phi_E,\phi_N)$ and transmission efficiency of a GMI vs. the scanning phase, which, here, is the roundtrip phase of the eastern arm, $\phi_E$. Each trace corresponds to a different bias point of the roundtrip phase in the northern interferometer arm, $\phi_N$. As the biasing phase is scanned, the $\gamma$ phase approaches a step function with respect to scanning phase, and the transmission exhibits narrowing peaks at $\phi_E=2\pi-\phi_N$. }
\end{figure}
It is important to note that, in the presence of loss, there is a trade-off between the phase resolution and the output power at the GMI transmission port. At bias points of increasing sensitivity, the transmitted power drops exponentially as a result of accumulated loss in the interferometer cavities. This effect is depicted in Figure \ref{fig:transloss} below, where the plot in Fig. \ref{fig:transloss}a shows the GMI behavior for lossless mirrors at several bias points, and the plot Fig. \ref{fig:transloss}b shows the same for mirror transmission and loss of 15x10$^{-6}$ and 5x10$^{-6}$, respectively. Although the reduction in transmitted power at the detector presents a challenge in locating the bright fringe, it allows the interferometer to operate purely on the bright fringe without the need to filter out residual pump light, as it is already weak.
\begin{figure}[ht!]
\centering
\includegraphics[width=\linewidth]{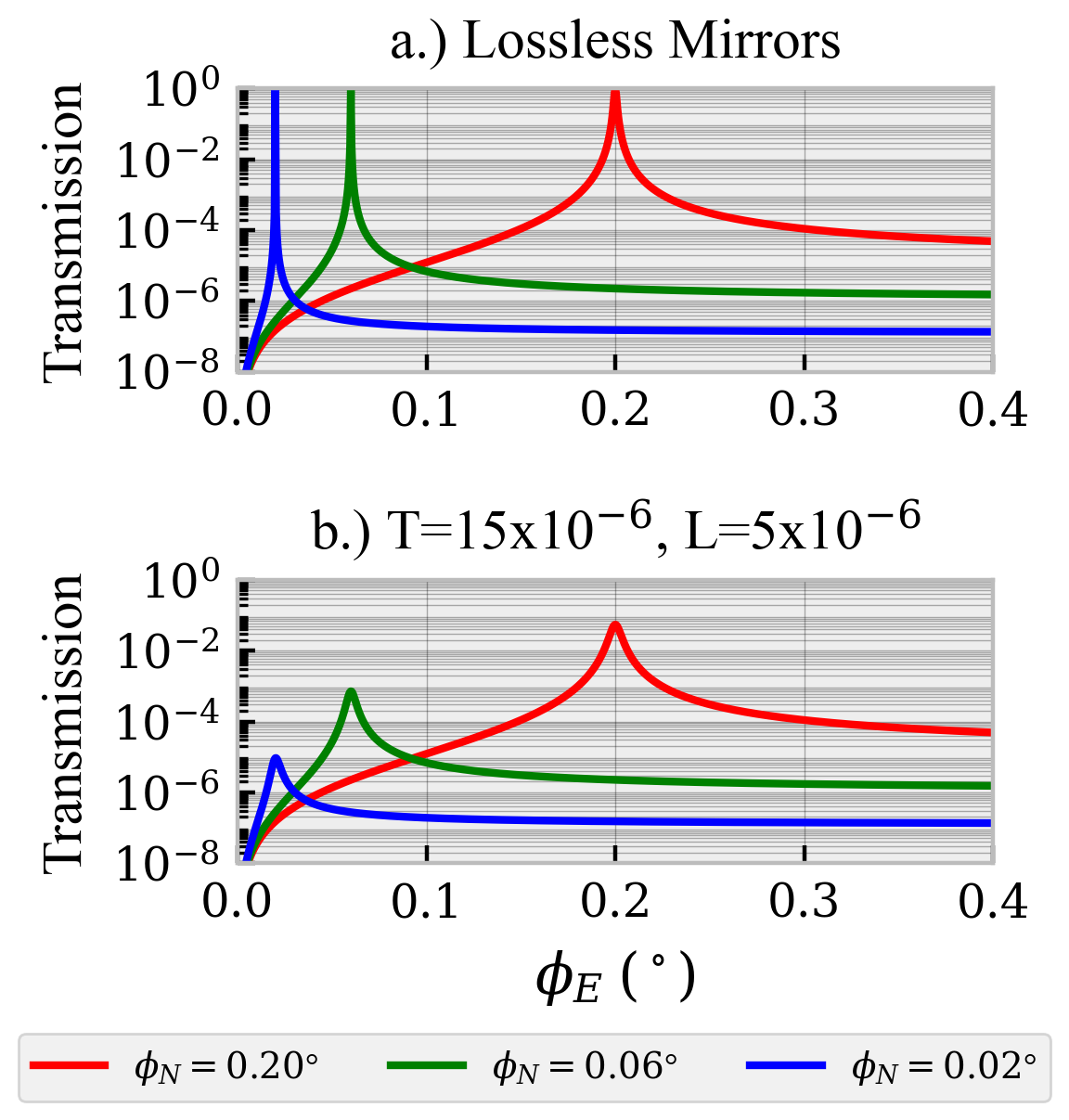}
\caption{\label{fig:transloss} Normalized GMI transmission vs. eastern arm phase $\phi_E$ at various $\phi_N$ bias points in the (a) absence and (b) presence of loss in all of the mirrors. Mirror loss has the effect of reducing transmission at the bright fringe and broadening the peak with respect to the scanning phase.}
\end{figure}
\vspace{-1em}
\subsection{Conventional light recycling}
\begin{figure}[ht]
\centering
\includegraphics[width=.9\linewidth]{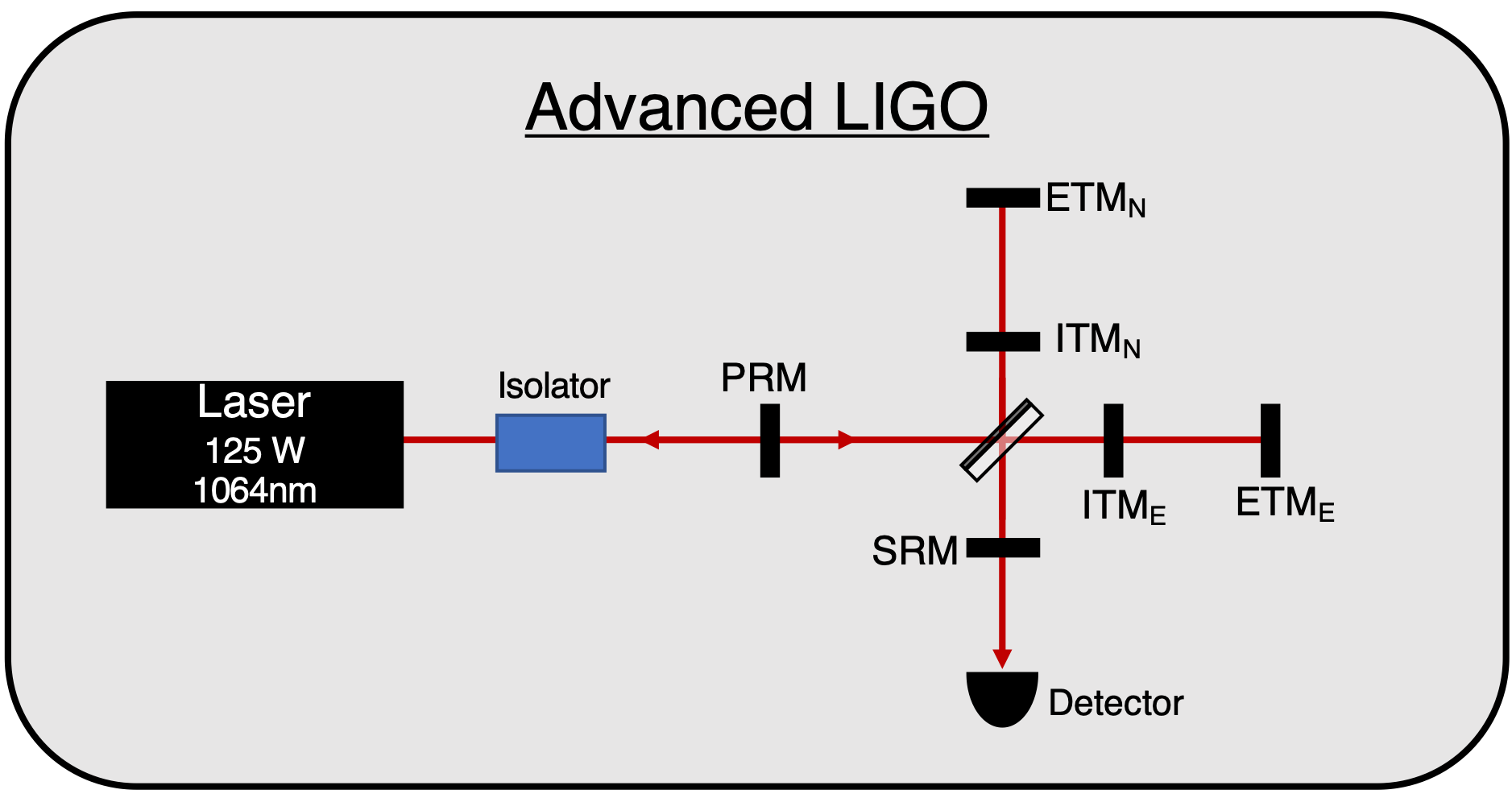}
\caption{\label{fig:ligo} Optical layout diagram for the simplified aLIGO configuration evaluated in our numerical study. Power recycling mirror (PRM) and signal recycling mirror (SRM) have power transmittances of 3\% and 20\%, respectively. End test masses (ETMs) are set to a relative phase of $\pi-0.00025^\circ$ to bias at the dark fringe. The internal test masses (ITMs) have a power transmittance of 1.4\%, and the interferometer arms are 4km long.}
\end{figure}

Advanced LIGO implements a dual recycled Fabry-Perot Michelson interferometer, in which a 125 W laser operating at 1064 nm is injected through the back side of a power recycling mirror (PRM) and into the central beam splitter \cite{aligo}. Each arm has a Fabry-Perot cavity consisting of suspended mirrors, the internal test mass (ITM) and end test mass (ETM). A signal recycling mirror (SRM) is placed at the antisymmetric (dark) port of the interferometer. 
\\The Michelson interferometer is operated at a small DC bias from the dark fringe, rendering the MI portion a single effective mirror with reflectivity proportional to the differential arm length. The MI mirror and PRM form a Fabry-Perot cavity, commonly referred to as the power recycling cavity (PRC), which allows laser power to build up and amplify the sensitivity of the device to GW-induced perturbations in the test masses \cite{Fritschel:powerrec}. The Fabry-Perot cavities increase the effective path length, recirculating power, and storage time of light in the arms. Proper control of the SRM position allows a single sideband frequency to resonate between the MI output and the SRM, increasing the signal power by constructive interference \cite{Meers:recycling}. The optical layout diagram for the aLIGO benchmark considered in this study is provided in Figure \ref{fig:ligo}. 

\section{Signal Sidebands in a Grover-Michelson Interferometer}
Assuming a simple sinusoidal modulation due to an $h_+$-polarized GW, the phase accumulated by light along the path of a harmonic spacetime distortion is given in Ref. \cite{Meers:recycling} as
\begin{equation}
    \phi = k_0L \mp \frac{\omega_0}{2}\int_{t-L/c}^{t} h_0\cos(\omega_{gw}t+\phi_{gw})dt = k_0L \mp \delta\phi \;,
\end{equation}
where $\delta\phi$ is the phase modulation,
\begin{equation}
   \delta\phi = \frac{h_0\omega_0}{\omega_{gw}}\cos\left(\omega_{gw}t+\phi_{gw}-\omega_{gw}\frac{L}{c}\right)\sin\left(\omega_{gw}\frac{L}{2c}\right)\;,
\end{equation}
$h_0$, $\omega_{gw}$, and $\phi_{gw}$ are the gravitational wave strain magnitude, frequency, and phase, respectively. Note that strain, $h=\delta L/L$, is a unitless quantity. Phase modulation induces new sideband frequencies with wavenumber $k_{gw}=\omega_{gw}/c$ that exit the interferometer and can be detected by operating at a small offset from the dark fringe and demodulating the photodetector beat signal at the gravitational wave frequency \cite{finesse}. The practice of operating at a small offset, or ``DC bias,'' from the dark fringe is referred to as DC readout \cite{Fricke:dcreadout} and will be considered exclusively throughout this paper. Alternative approaches that bias the device purely on the dark fringe include homodyne readout using a local oscillator that beats together with the interferometer output, or heterodyne readout, which involves modulating the input carrier before the interferometer to generate a high frequency sideband that mixes with GW-generated sidebands \cite{chen:heterodyne}.
\\In the case of a GMI operating at a point of strong cavity coupling, the carrier field in the arms and the output amplitudes of the sidebands are amplified by the magnitude of the average recirculating cavity power. This increases the effective number of times the carrier wave bounces off the end mirrors, similar to the effect of adding a power recycling mirror to a conventional MI, and it is variable by tuning the bias phases. The transfer function from a gravitational wave signal to the output photodiode signal in a GMI, $T_{\text{GM}}(\omega_{gw})$, is the same as that of the MI multiplied by the cavity field magnitude (c.f. Eq. 5.41 in ref \cite{finesse}),
\begin{equation}\label{eq:agw}
 T_{\text{GM}}(\omega_{gw}) = \frac{P_0|E_c|\omega_0}{\omega_{gw}}\sin(k_0\delta_{\text{off}})\sin\left(\frac{\omega_{gw}\bar{L}}{c}\right)\left[\frac{\SI{}{\watt}}{\text{h}}\right]
\end{equation}
where $\bar{L}=(L_N+L_E)/2$ is the common arm length. Note that gravitational waves with frequencies that are integer multiples of $2\pi c/\bar{L}$ interfere destructively in the interferometer due to the quadrupolar nature of the radiation. The field $E_c=(E_{N}-E_{E})/2E_0$ is the average of the two cavity relative field amplitudes; the minus sign arises because in each cavity one of the recirculating fields reflects off the Grover coin back into the same cavity while the other cavity field transmits through it. The power at the detector that is linear in $h$ is also linear in both the input laser power and the magnitude of the cavity field enhancement \cite{yuen3}. Assuming unit reflectance in mirror M2, roundtrip phase $\phi_2=0$, and identical test-mass mirrors with amplitude reflectance $r=\sqrt{R}$, the power enhancement gain factor, $\mathcal{G}$, for a given bias point $\phi_N=-\phi_E=\pm\theta$ can be written as
\begin{equation}
    \mathcal{G}(\theta)=\frac{R\sin^2\theta}{2+2R\cos^2 \theta -4R\cos \theta }\;.
\end{equation}
In principle, $\mathcal{G}$ depends on $\omega_{gw}$ since $\phi_{N,E}=2k_0L_{N,E}\pm\delta\phi$ where $\delta\phi\propto\omega_{gw}$, but the strain $h$ is on the order of $10^{-21}$ which has negligible effect on the phase of the carrier wave, so $\mathcal{G}$ is a broadband quantity similar to the gain factor in power recycling.
\\The minimum detectable strain amplitude of a gravitational wave is quantified by the NSR of the sideband amplitudes exiting the interferometer. The shot-noise limited NSR for the GMI is (c.f. Eq. 6.21 in ref \cite{finesse})
\begin{equation}\label{eq:nsr}
    \text{NSR}_{\text{GM}} = \sqrt{\frac{2\hbar}{\omega_0P_0\mathcal{G}}}\frac{\omega_{gw}}{\sin(\omega_{gw}\bar{L}/c)} \left[ \frac{\text{h}}{\sqrt{\SI{}{\hertz}}}\right]
\end{equation}
which is a factor of $1/\sqrt{\mathcal{G}}$ smaller than that of a conventional Michelson interferometer. Note that in the presence of mirror loss, a full treatment in the two-photon formalism, established by Caves and Schumaker \cite{caves:qmnoise,caves:tpf1,caves:tpf2} and applied to GW detectors by \textcite{chen:gwqnoise,chen:tpfmath}, would be required. A complete derivation of the NSR of a lossy GMI is beyond the scope of this work, but the method is implemented in the \textsc{finesse} software. The solid lines in Figure \ref{fig:micomp} below show the results of equations \eqref{eq:agw} and \eqref{eq:nsr} compared to their counterparts for a standard MI (Eq 5.41 and 6.21 in ref \cite{finesse}). 
\begin{figure}[ht]
\centering
\includegraphics[width=.95\linewidth]{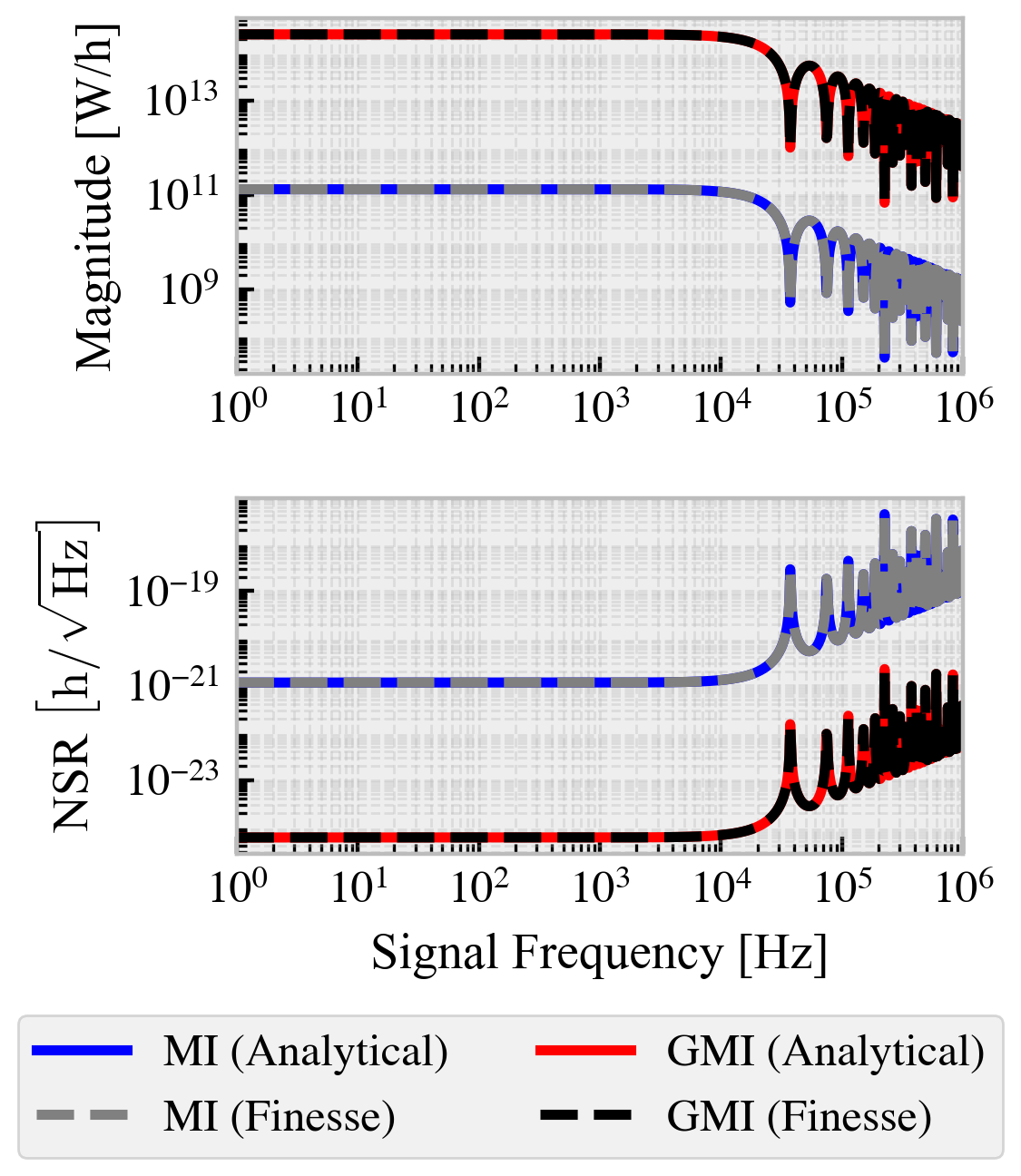}
\caption{\label{fig:micomp} Magnitude (top) and noise-to-signal ratio, or strain sensitivity (bottom), of the GW transfer function amplitude for a Grover-Michelson and a Michelson interferometer with 4km arm lengths, input power of 125 W, and lossless, stationary end mirrors. In each scenario, a small DC offset bias on the order of 1mrad is employed to allow signal extraction. Solid red lines indicate the analytical results of equations \eqref{eq:agw} and \eqref{eq:nsr} for the GMI biased at $\phi_E=-\phi_N=0.06^\circ$.}
\end{figure}
\\The GMI was biased at a round-trip phase of $\phi_N=-\phi_E=0.06^\circ$, which was empirically found to provide strong sensitivity and bandwidth in the GMI. In this configuration, the Grover-Michelson interferometer is more than three orders of magnitude more sensitive than a standard Michelson interferometer without light recycling. To confirm the validity of the analytical equations, we simulated the GMI using the \textsc{python} interface for the open source interferometer simulation program \textsc{finesse} (frequency-domain interferometer simulation software). The software calculates noise by propagating the quantum vacuum field from all the open ports in the interferometer to the readout using the transfer matrix method \cite{finesseref}. The numerical data are shown by the dashed lines and agree well with the analytical values. 
\section{Numerical Comparison of Grover-Michelson Interferometer with aLIGO}
Due to the extremely small strain amplitudes associated with GW radiation, mechanical noise in suspended test masses dominates the sensitivity spectrum at levels beyond the shot-noise limit. 
Radiation pressure noise is particularly strong at low frequencies and is proportional to the power recirculating in the arm cavities and the masses of the suspended mirrors \cite{finesse}. 
To quantify this effect, we simulated the GMI in \textsc{finesse} to compare it with the advanced LIGO interferometer as a baseline. We used the parameters in the \textsc{finesse} example file for aLIGO \cite{finesse} for both interferometers, which include a 1.064µm laser power of 125 W and 40kg test masses with 1-million Q resonances at 1 Hz. 

\begin{figure}[ht]
\centering
\subfigure{\includegraphics[width=.9\linewidth]{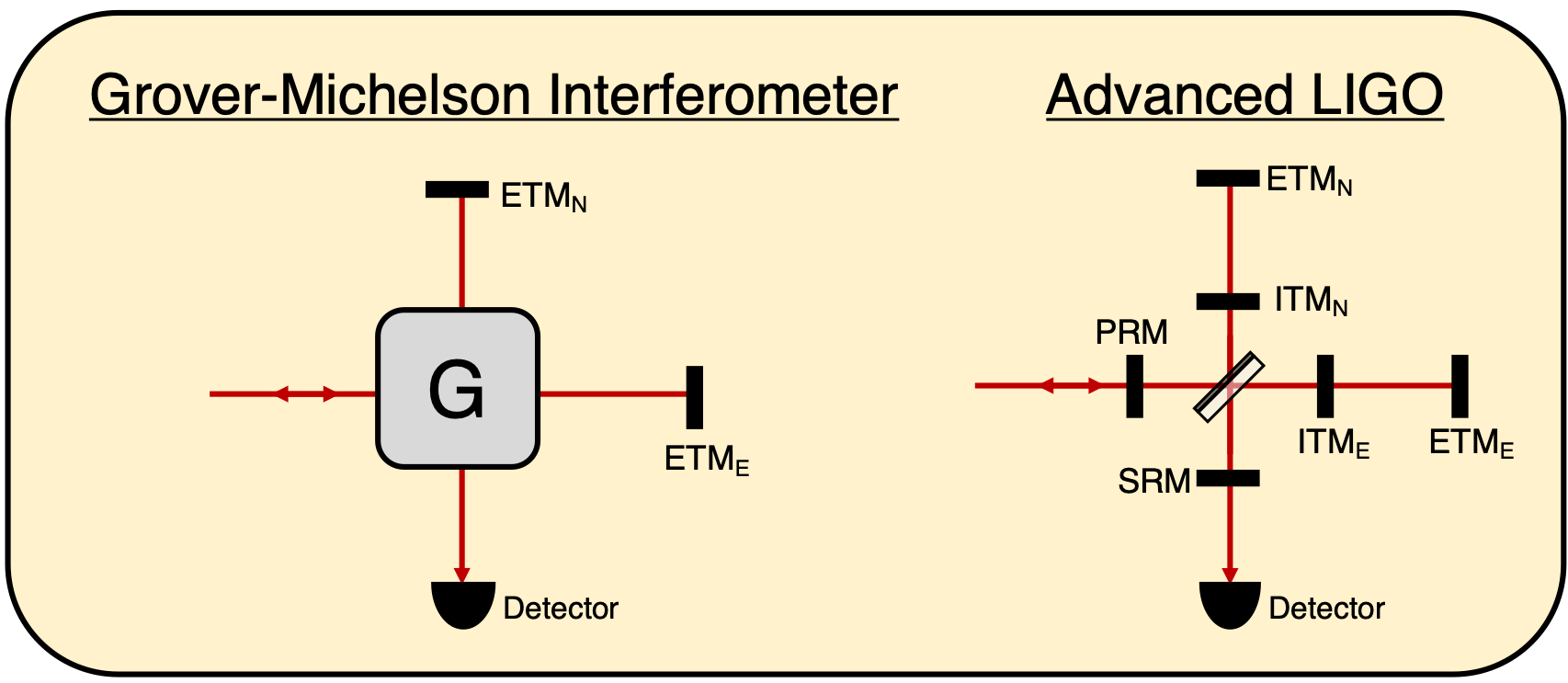}}
\subfigure{\includegraphics[width=\linewidth]{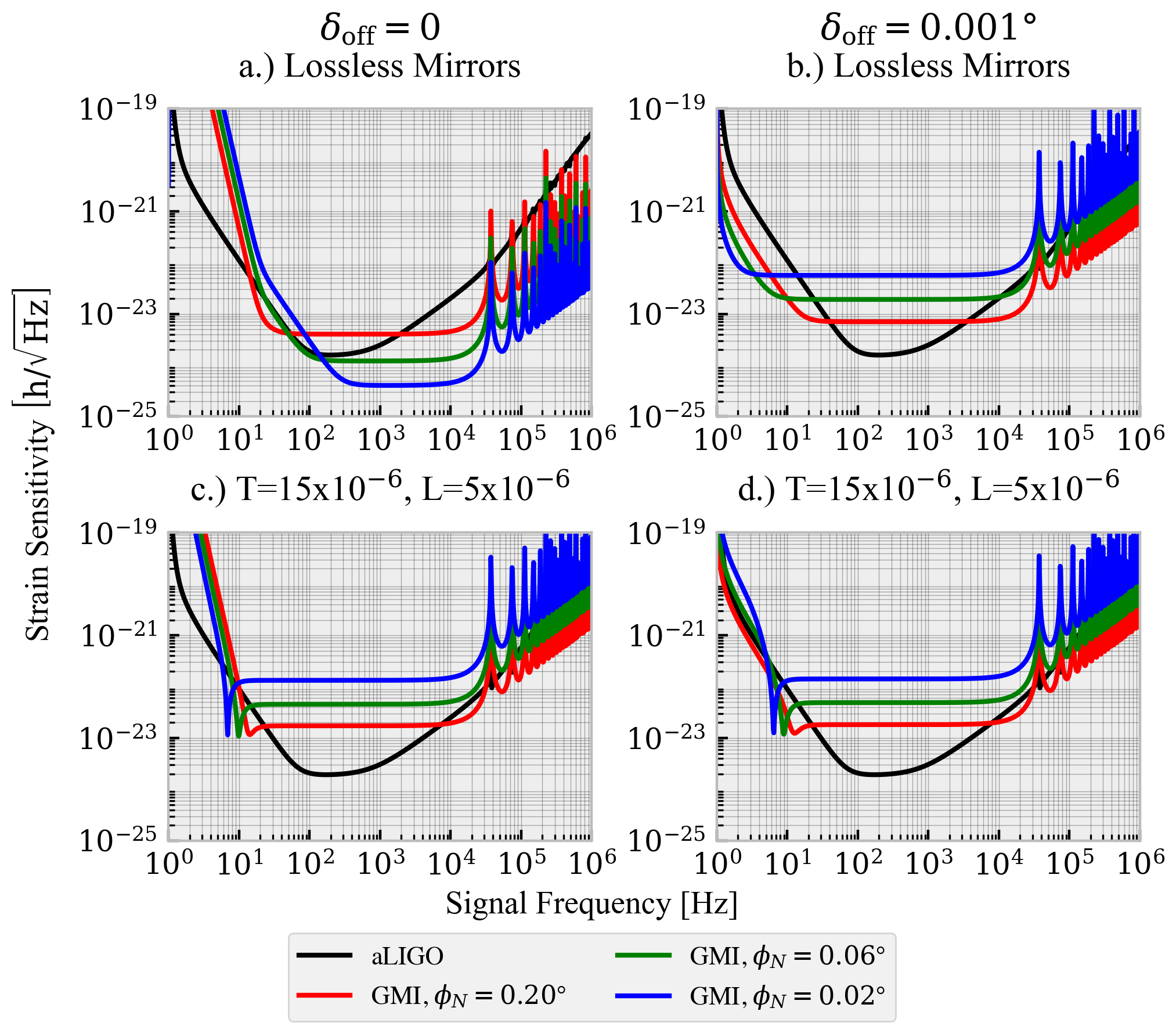}}
\caption{\label{fig:ligocomp} (Top) Optical layout diagrams of the Grover-Michelson interferometer and simplified aLIGO configuration. (Bottom) Comparison of strain sensitivity in the presence of radiation pressure noise for aLIGO and various settings of the GMI bias point (a, b) without mirror losses and (c, d) with 15x10$^{-6}$ power transmission and 5x10$^{-6}$ power loss in the mirrors. For each GMI configuration, $\phi_E$ is always biased to the point $2\pi-\phi_N$. In all cases, we consider lossless, balanced beam splitters, a laser power of 125 W, and radiation pressure noise effects.}
\end{figure}
In all the GMI settings considered, the phases are such that $\phi_E=-\phi_E-\delta_{\text{off}}$, where $\delta_{\text{off}}$ is a DC offset bias. In the left column of Figure \ref{fig:ligocomp}, we set $\delta=0^\circ$, and in the right column we set $\delta=0.001^\circ$. In both the lossless and lossy cases, the $\phi_N$ bias point provides a tunable response with enhancements over the aLIGO benchmark. With lossless mirrors, the GMI can provide a broadband enhancement of over an order of magnitude at higher frequencies with zero offset bias and lower frequencies with a small offset. This behavior is reminiscent of tuned recycling \cite{Meers:recycling}. With lossy mirrors, the GMI can provide narrowband enhancements below 30Hz and a tunable notch that narrows as the bias point approaches $(\phi_N,\phi_E)=(0,0)$, and the small offset has a negligible effect. The reason for the reduced sensitivity in the GMI compared to aLIGO in the presence of mirror transmission is due to the fact that the lossy GMI has more open ports through which vacuum fluctuations can enter and contribute to noise. In principle, a single-component Grover coin would not experience this effect, as it would have the same number of open ports as aLIGO, and perhaps be more robust to loss in the end mirror coatings.
\section{Enhancing GMI Performance further with Light Recycling Configurations}
\begin{figure}[ht!]
  \centering
  \includegraphics[width=\linewidth]{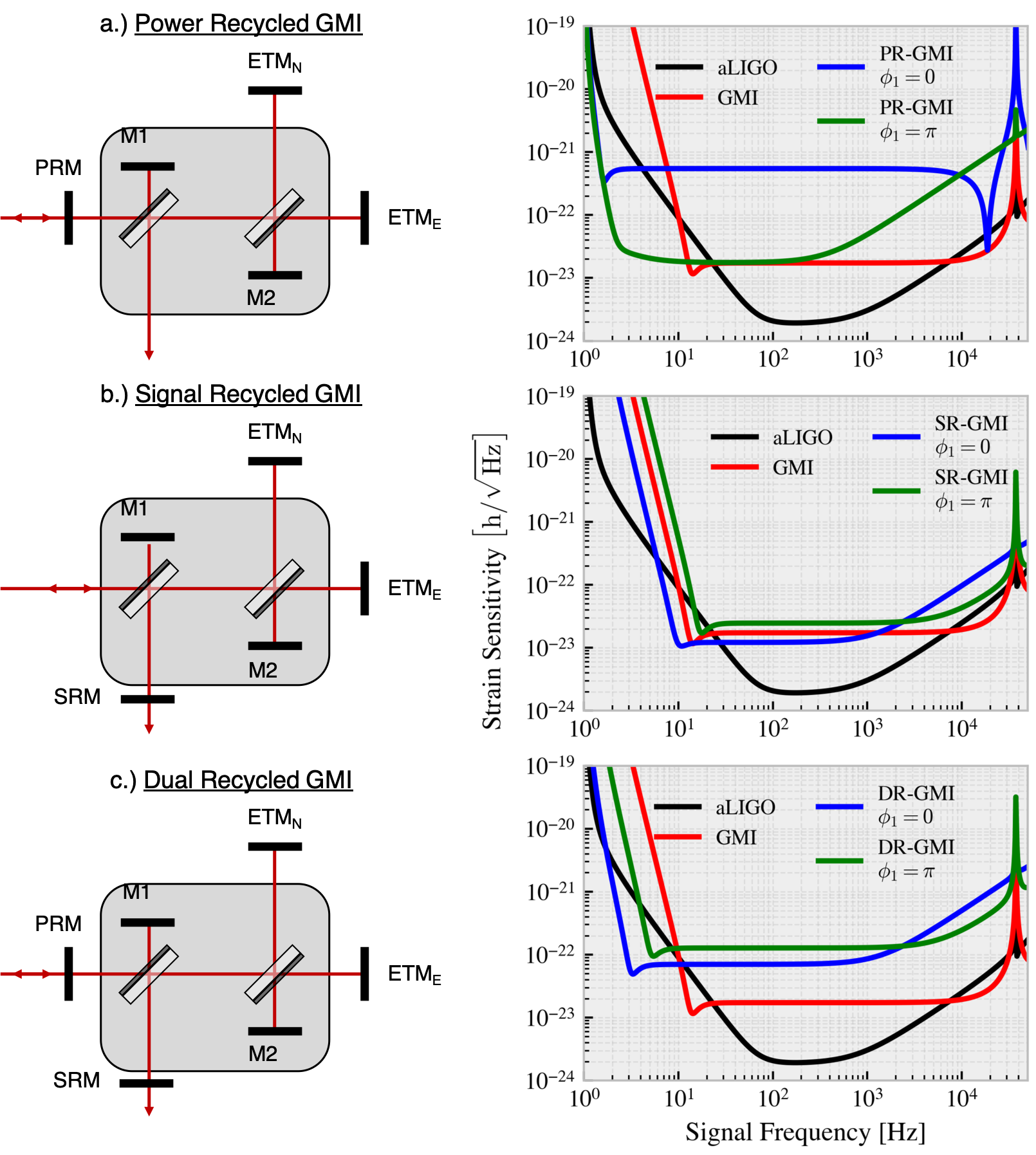}
  \caption{\label{fig:recycle} Optical layout diagrams and numerically simulated strain sensitivity spectra of (a) Power Recycled GMI, (b) Signal Recycled GMI, and (c) Dual Recycled GMI. Standard GMI and aLIGO performances are shown in red and black on each plot, respectively. Each interferometer was simulated with internal Grover coin phase $\phi_1$ set to 0$^\circ$ and 180$^\circ$, displayed in blue and green, respectively. Radiation pressure is taken into account for each trace, and each mirror possesses a power loss of 5x10$^{-6}$. The power transmission of the PRM is 0.1\% for the PR configuration and 3\% for the DR configuration. The SRM power transmission is 20\% in both the SR and DR devices. The GMI bias point in each case is ($\phi_N,\phi_E)=(0.2^\circ,-0.2^\circ)$.}
\end{figure}

To evaluate whether GMI performance could be improved, we investigated its combination with several light recycling schemes numerically using \textsc{finesse}. Optical layout diagrams and corresponding strain sensitivities for power recycling (PR), signal recycling (SR) and dual recycling (DR) in a GMI are shown in Figure \ref{fig:recycle}. We implemented the same laser power, mirror parameters, and radiation pressure effects as in Figure \ref{fig:ligocomp}, with $\phi_E=-\phi_N=0.2^\circ$. 
\\The PRM power transmission was set to 0.1\% in the PR configuration to maximize low-frequency enhancement, and set to 3\% in the DR configuration to match the PRM transmission in the aLIGO configuration. The SRM power transmission was set to 20\% to match that of the aLIGO configuration. In each plot, the standard GMI and aLIGO benchmark are included for comparison. We also show the effect of switching the first Grover coin internal mirror M1 between relative phase shifts of $\phi_1=0^\circ$ and 180$^\circ$, i.e., between transmitting and reflecting behavior. 

The addition of a power recycling mirror to the GMI provides a quantum-limited strain sensitivity of less than $3\times10^{-23}[\text{h}/\sqrt{\SI{}{\hertz}}]$ at frequencies in the range $\sim$1-300 Hz, which is an improvement over the aLIGO configuration at frequencies below 20 Hz by up to two orders of magnitude. Signal recycling provides a slight improvement over the standard GMI at frequencies below 1kHz. For dual recycling, there is only a modest improvement in $10^{-22}[\text{h}/\sqrt{\SI{}{\hertz}}]$ at frequencies between 1 and 20 Hz compared to both aLIGO and the standard GMI, although it is possible that we did not consider optimal recycling mirror parameters such as transmittance and phase. 
\\In the case of power recycling, signal enhancement occurs when the GMI is in reflective mode ($\phi_1=180^\circ$), establishing a Fabry-Perot cavity between the PRM and the GMI, as in conventional power recycling. In the SR configuration, this enhancement occurs when the GMI is highly transmissive to the carrier (left) input and reflective to the recycled (bottom) input, thereby establishing a Fabry-Perot cavity between the SRM and the GMI. Reduction in sensitivity at high frequencies is likely the result of the storage time in the nested cavities exceeding the finite coherence time of the pump laser.
\section{Practical considerations for a long-baseline GMI}
Until now, only a benchtop implementation of the GMI has been demonstrated \cite{Schwarze:gmi_exp}, raising valid questions about its viability as a long-baseline GW detector. State-of-the-art mirror positioning and feedback are implemented in modern GW detectors to ensure that the cavity test masses are controlled within 0.1pm RMS, and beam pointing is controlled within approximately 10nrad RMS to avoid coupling to higher-order transverse modes \cite{LIGO_control,LIGO_daq,LIGO_ctrldata,LIGO_ctrlsystem,LIGO_controlmodel}. This level of control would be sufficient to implement a GMI-based detector with a bias point of 0.001$^\circ$ relative phase, or about 3 pm. The mirrors in a light-recycled MI typically have a slight curvature to mimimize transverse mode coupling, so a GMI would require the same for all mirrors in the device.
\\ The simplified aLIGO configuration considered here requires positioning of the PRM, SRM, beam splitter, and two suspended mirrors in each cavity, or seven components. A GMI would require positioning M1, M2, both beam splitters, and the two suspended test masses, or six components, with an optional seventh if a PRM or SRM is implemented. In practice, all major components of aLIGO are suspended to isolate them from seismic and thermal noise. The same case would apply to a GMI, though the entire Grover coin assembly could be mounted on the same suspension platform, as if it were a single component.
\\With current state-of-the-art mirror coatings \cite{LIGO_coatings}, a scaled version of the GMI could provide enhanced sensitivity in the sub-50 Hz range, especially if a PRM is used. To enable the full range of tunable sensitivity enhancement, i.e., the data displayed in the top row of Fig. \ref{fig:ligocomp}, mirror transmission and loss on the order of $10^{-8}$ would be required. Although this is outside the current capabilities of most coating vendors, the field is constantly reducing loss through advancement in materials science and fabrication methods.
\\In this study, we did not consider squeezed vacuum inputs, although it is straightforward to analyze numerically and provides a reduction in quantum shot noise, regardless of the layout of the interferometer \cite{caves:qmnoise}. 
\\An important feature for the practical implementation of a GMI for GW detection is that, for a given $\phi_N$ bias, setting $\phi_E=-\phi_N$ places the interferometer on a \textit{ bright fringe}. Ideally, the intensity of the carrier wave at the output is small to avoid detector saturation, which is why the interferometer is conventionally biased a small way off the \textit{dark fringe}. For a small $\phi_N$ bias in a GMI, the transmission drops off exponentially from the bright fringe, so a small DC offset in $\phi_E$ from this point also provides low carrier intensity at the detector while maintaining strong sensitivity to path-length distortions. Alternatively, the GMI transmission behavior can be converted to reflection by biasing the first internal mirror M1 to a point of round-trip phase $\phi_1=\pi$ so that the device can be biased off a \textit{dark fringe}. Alternative approaches to removing unwanted strong carrier light include the use of a Fabry-Perot cavity at the output of the device \cite{Miao:qlimits}.
\subsection{Laser Frequency and Intensity Noise}
\begin{figure}[ht!]
  \centering
  \includegraphics[width=\linewidth]{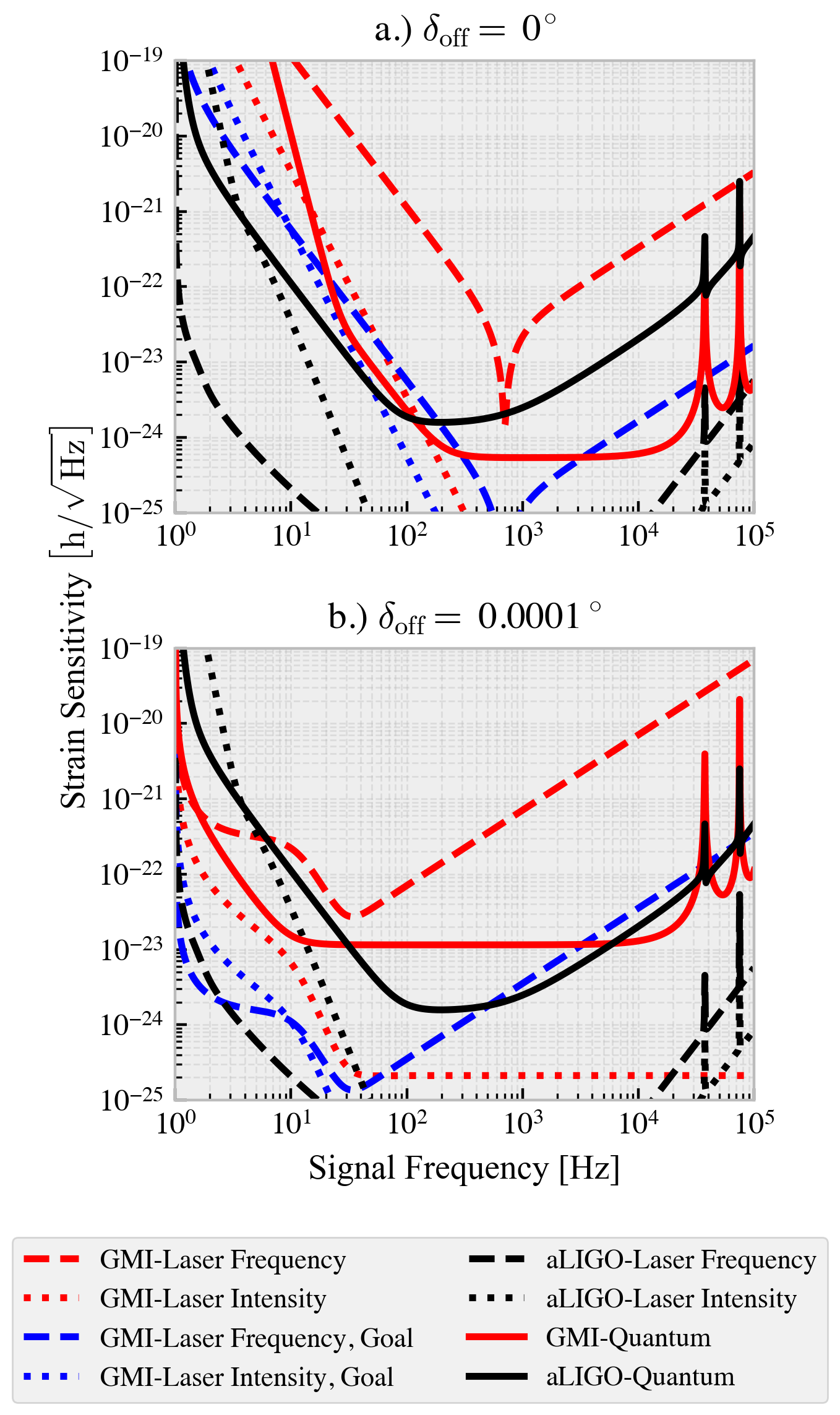}
  \caption{\label{fig:lasernoise} Noise budget for practical and targeted out-of-loop laser source noise in a GMI GWD. GMI bias points used to generate data were a.) $(\phi_N,\phi_E)=(0.02^\circ,-0.02^\circ)$ and b.) $(0.02^\circ,-0.0201^\circ)$. Coating transmission and loss values were set to $15\times10^{-9}$ and $5\times10^{-9}$, respectively, to demonstrate performance limit.}
\end{figure}

As a result of the GMI performance strongly depending on both the common and differential arm lengths, it does not possess the same common-mode suppression as conventional MIs do. Thus, a GMI-based GWD would require a laser source with reduced intensity and frequency noise beyond what is achieved in aLIGO's pre-stabilized laser (PSL) system, which involves three stages of optical feedback and control loops \cite{LIGO:PSL}. The final control stage of the PSL in aLIGO involves locking the laser frequency to the 4km arm length using in- and out-of-loop detectors. A similar control scheme for the GMI is envisioned where the in-loop detection occurs between the two beam splitters of the Grover Coin.
\\Laser noise refers to fluctuations in the instantaneous frequency or intensity of the carrier laser, which can appear as differential phase fluctuations in the arms and overwhelm the gravitational-wave signal. The out-of-loop laser frequency and intensity noise, i.e. the remnant noise propagating through the interferometer after the control loops, are on the order of $10^{-6}\,\SI{}{\hertz}/{\sqrt{\SI{}{\hertz}}}$ and $10^{-9}\,\SI{}{\watt}/\sqrt{\SI{}{\hertz}}$ at 100 Hz, respectively, according to data published from aLIGO's third and fourth observing runs \cite{LIGO:noisedata2020,aLIGO:4thObs}. To compute the laser frequency and intensity noise budgets in a GMI GWD, the transfer function was calculated for each, multiplied by their reported power spectral densities (PSDs), and divided by the gravitational wave signal to obtain units of strain$/\sqrt{\SI{}{\hertz}}$.
\\Figure \ref{fig:lasernoise} displays the contributions of laser and frequency noise to the strain sensitivity spectrum of a GMI GWD compared to those of aLIGO. Laser frequency noise is represented by dashed lines, intensity noise is shown by dotted lines, and the quantum limit including radiation pressure is shown by solid lines. The operating points of the GMI considered here are ($\phi_N$, $\phi_E$) = ($0.02^\circ$, -$0.02^\circ-\delta_{\text{off}}$), where the offset $\delta_{\text{off}}$ is $0^\circ$ in Fig. \ref{fig:lasernoise}a and $0.0001^\circ$ in Fig. \ref{fig:lasernoise}b, to demonstrate the effect of laser noise in the high- and low-frequency operating modes of the GMI, respectively.
\\The red dashed and dotted lines represent laser noise contributions from the reported out-of-loop values of aLIGO. The blue traces represent noise generated by reduced out-of-loop noise values that would enable enhancements in GW strain sensitivity in the GMI. The out-of-loop laser frequency noise required to enable GW detection in a GMI is two orders of magnitude lower than that published in the third and fourth aLIGO observation runs. The out-of-loop intensity noise required to achieve the same effect is a factor of six lower than that reported. Obviously, such a reduction in laser frequency noise would require considerable advancement in laser stabilization technology, which presents the greatest hurdle to enabling the use of a GMI as a long-baseline gravitational wave detector.
\subsection{Coating Thermal Noise}
\begin{figure}[ht!]
  \centering
  \includegraphics[width=\linewidth]{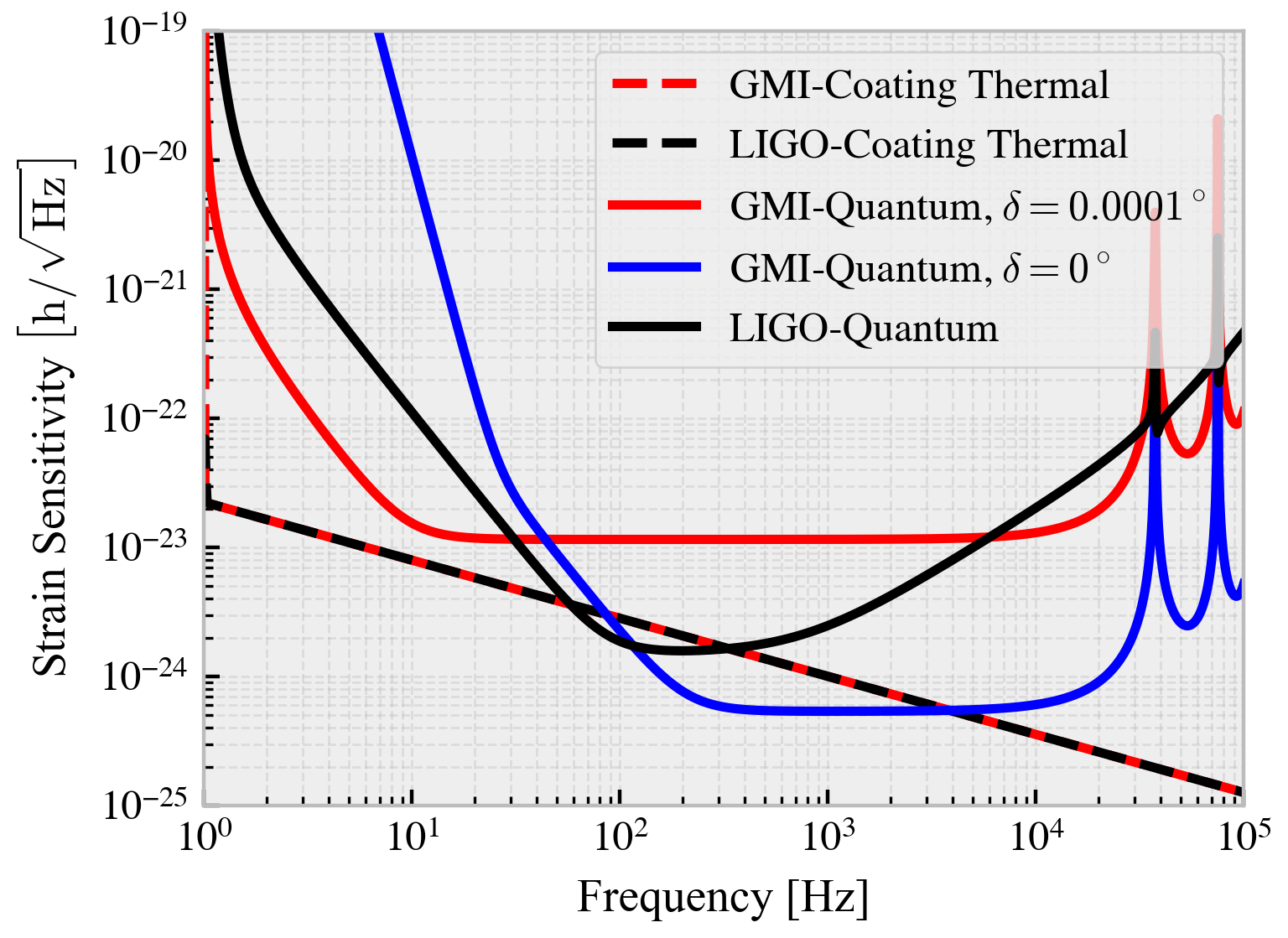}
  \caption{\label{fig:coatingnoise} Noise budget for total coating thermal noise in a GMI GWD compared with that of aLIGO. The reported CTN spectrum from aLIGO's fourth observing run was used as input. Data for the GMI were generated using a bias point of ($\phi_N$, $\phi_E$)=(0.02$^\circ$, -0.02$^\circ-\delta_{\text{off}}$), where $\delta_{\text{off}}=0^\circ$ in a.) and $0.0001^\circ$ in b.).}
\end{figure}

Another source of noise in GW detectors is the random thermal motion in the mirror suspensions, substrates, and optical coatings. The coating thermal noise due to Brownian motion is particularly dominant at low frequencies and eclipses the quantum noise of aLIGO near 100 Hz. The mirrors coatings are based on Bragg reflectors, comprised of many layers of alternating high and low refractive index, to create highly reflective surfaces. To quantify the effect of CTN in the GMI, we used the PSD of total thermal noise, including substrate and coating noise reported in ref \cite{aLIGO:CTN}, which was estimated to be
\begin{equation}
    \sqrt{S_{\text{CTN}}}=1.13\times10^{-20}\left(\frac{100 \SI{}{\hertz}}{f}\right)^{0.45} \frac{\SI{}{\meter}}{\sqrt{\SI{}{\hertz}}}\,.
\end{equation}
This is multiplied by the ratio of transfer functions between the single mirror motion and differential mirror motion in each interferometer, which is essentially unity at all frequencies. To obtain units of strain, we divide by the cavity length of 4 km. The results of CTN in a GMI GWD and aLIGO are presented in Figure \ref{fig:coatingnoise}, where the coating noise is in dashed lines and the quantum noise is in solid lines. The same GMI bias points were used as in Figure \ref{fig:lasernoise}, and coating transmission and loss values were set to $15\times10^{-9}$ and $5\times10^{-9}$, respectively. The data suggest that the CTN is sufficiently subdominant to enable the low-frequency sensitivity enhancement in the GMI, but is dominant at frequencies below 3 kHz in the high-frequency GMI setting. At frequencies above 300 Hz, the total sensitivity can still provide an advantage over aLIGO.
\subsection{Power Handling Considerations}
Due to the high recirculating powers in interferometers such as the GMI and dual-recycled FP MIs, thin-film coating damage and thermal lensing must be avoided to maintain device uptime and stable operation. Table \ref{table:powers} provides the simulated steady-state power levels incident on critical components in the GMI and aLIGO. The GMI bias point used to generate the data was the same as in the previous three figures. The powers at both internal and external cavity mirrors in aLIGO are the same for the northern and eastern arms, so the entries in the table are simply labeled ``ITM'' and ``ETM'' for brevity. 
\begin{table}[h!]
    \centering
    \caption{\label{table:powers} Steady-state power levels incident on critical components in a GMI and aLIGO for 125 W of input laser power, calculated using power detectors in \textsc{finesse}. The GMI bias point is $(\phi_N,\phi_E)=(0.02^\circ,-0.0201^\circ)$. A small amount of loss on the order of 10$^{-8}$ is applied to each of the components in both the GMI and aLIGO, representing ``lossless'' performance without allowing the calculated cavity powers to approach unrealistic values.}
    \setlength{\tabcolsep}{5pt} 
    \renewcommand{\arraystretch}{1.5} 
    \vspace{1em} Grover-Michelson \\ \vspace{1em}
    \begin{tabular}{|c|c|c|c|}
    \hline
    \textbf{BS2} & \textbf{M2} & \textbf{ETM}$_{\text{N}}$ & \textbf{ETM}$_{\text{E}}$ \\
    \hline
    2.52 MW & 2.52 MW & 1.25 MW & 1.27 MW \\
    \hline
    \end{tabular}
    \\
    \vspace{1em}
    Advanced LIGO
    \\
    \vspace{1em}
    \setlength{\tabcolsep}{6pt} 
    \begin{tabular}{|c|c|c|c|}
    \hline
    \textbf{BS} & \textbf{PRM} & \textbf{ITM} & \textbf{ETM} \\
    \hline
    16.4 kW & 16.4 kW & 2.33 MW & 2.33 MW \\
    \hline
    \end{tabular}
\end{table}
\\The recirculating power in the Fabry-Perot arm cavities of aLIGO is on the order of $2\times10^{6}$ watts, or 2 MW. In the GMI, the recirculating power increases as the round trip phase in both arms approaches $0\mod{2\pi}$, and is unbounded in principle, provided that an infinite number of round trips is obtained. For example, monitoring the power incident on the Grover coin mirror M2 in a \textsc{finesse} simulation yields a value of 200 MW for a GMI set to the bias point used in Figures \ref{fig:lasernoise} and \ref{fig:coatingnoise}. Note that, in such a simulation, only plane waves are considered. 
\\In practice, however, transverse mode coupling and loss in the optical components will limit the achievable number of round trips. Setting the coating loss to levels of $10^{-8}$ for all components in the GMI allows the same GW sensitivity as with lossless components, but the recirculating power drops to the same 2 MW level as in the FP arm cavities of aLIGO. 
\\To avoid coating damage and minimize thermal lensing in aLIGO, thin-film coatings are deposited by ion-beam sputtering, which is known to produce dense films with low absorption loss, high laser damage thresholds, and low thermal noise \cite{aLIGO:CTN}. Large beams are also used to minimize power density and reduce Brownian noise in the coatings \cite{aLIGO:CTN}. The same protocols would be required for the GMI, and a larger beam radius may be needed to ensure that the high power at the second Grover coin beam splitter, BS2, does not cause considerable thermal lensing.
\section{Conclusion}
Through a physics-first approach, we derived the coupled cavity equations in a Grover-Michelson interferometer to determine the optimal bias points for achieving high recirculating power inside the interferometer. It has been shown through numerical simulation that the GMI has the potential to provide an advantage in strain sensitivity compared to aLIGO at frequencies below 30 Hz or in the 100Hz-10kHz range, which is tunable based on the biasing point. The range of low-frequency performance can be broadened by adding a power or signal recycling mirror and switching the GMI to reflective or transmissive behavior, respectively. The novelty of this device is that it has the potential to achieve state-of-the-art sensitivity to gravitational waves without the need for Fabry-Perot cavities in the arms, provided the laser source frequency and intensity noise can be sufficiently suppressed.
\\In the ongoing design of novel GW detectors, directionally unbiased multiports provide a new suite of tools to improve the sensitivity of these devices. The phase-tunability of the GMI can provide a new benefit compared to standard light recycling techniques, which require modification of the mirror coatings or variable reflectivity recycling mirrors to achieve modular behavior. As thin-film mirror coating technology progresses to the point of 10$^{-8}$ level loss figures, and laser noise is continually optimized, the GMI will provide a strong alternative to the current state-of-the-art in GW detection. Future work may be conducted to demonstrate the noise characteristics of a benchtop GMI system at higher modulation frequencies, along with analysis of different noise reduction and readout strategies.
\begin{acknowledgments}
This research was supported by the Air Force Office of Scientific Research MURI Award No. FA9550-22-1-0312.
\end{acknowledgments}
\section*{Data Availability Statement}
Codes for generating the data supporting the findings of this study are openly available in the repository ``GMI\_Finesse'' at https://www.github.com/anthmanni/GMI\_finesse, reference number \cite{GMIfinesse}.
\appendix
\section{Derivation of GMI cavity field amplitudes (\ref{eq:cavs})}
\label{app:cavs}
A Grover coin can be decomposed into two Y-couplers, each of which is composed of a beam splitter and a mirror placed at the reflective port \cite{Schwarze:yc}. The transfer matrix for a lossless, balanced dielectric beam splitter is
\begin{equation}
    \frac{1}{\sqrt{2}}\begin{pmatrix}
        1 & 1\\ 1 & -1
    \end{pmatrix} ,
\end{equation}
where the negative sign is the result of Fresnel reflectance at the air-film interface. The action of a mirror on an amplitude simply adds a coefficient of $-r$, or $re^{j\pi}$, where $r$ is the amplitude reflectivity of the mirror. Consider a field $E_0$ incident on the left port of the Grover coin in Fig. \ref{fig:graphic}. After being transmitted through the first beam splitter and traversing the space between the two beam splitters, the field is now $\tilde{E}_0=E_0e^{j\theta}/\sqrt{2}$. This amplitude impinges on the second beam splitter, routing it into the northern and eastern arms of the interferometer, reflecting off the mirrors back to the beam splitter. Depending on the round-trip phase in each arm, the amplitudes can recombine coherently to either exit the device or remain inside by reflecting back and forth between the end mirrors and the internal Grover coin mirror M2, or some combination of the two.
The fields in the cavities are coupled as follows:
\begin{equation}
    \begin{aligned}
    E_N=\frac{\tilde{E}_0}{\sqrt{2}}+\left(\frac{1}{\sqrt{2}}\right)^2(-r_N)(-r_2)e^{j(\phi_N+\phi_2)}E_N\\+\frac{1}{\sqrt{2}}\left(-\frac{1}{\sqrt{2}}\right)(-r_E)(-r_2)e^{j(\phi_E+\phi_2)}E_E\;,
    \end{aligned}
\end{equation}
\begin{equation}
    \begin{aligned}
    E_E=\frac{\tilde{E}_0}{\sqrt{2}}+\left(-\frac{1}{\sqrt{2}}\right)^2(-r_E)(-r_2)e^{j(\phi_E+\phi_2)}E_E\\+\frac{1}{\sqrt{2}}\left(-\frac{1}{\sqrt{2}}\right)(-r_N)(-r_2)e^{j(\phi_N+\phi_2)}E_N \;.
    \end{aligned}
\end{equation}
Collecting like terms results in \vspace{.5em}
\begin{equation}\label{eq:a1}
    \begin{aligned}
    E_N=\frac{E_0e^{j\theta}-r_Er_2e^{j(\phi_E+\phi_2)}}{2-r_Nr_2e^{j(\phi_N+\phi_2)}}\;,
    \end{aligned}
\end{equation}
\begin{equation}\label{eq:a2}
    \begin{aligned}
    E_E=\frac{E_0e^{j\theta}-r_Nr_2e^{j(\phi_N+\phi_2)}}{2-r_Er_2e^{j(\phi_E+\phi_2)}}\;.
    \end{aligned}
\end{equation}
\vspace{1em}Substituting \eqref{eq:a2} for $E_E$ in \eqref{eq:a1} and solving for $E_N$,\vspace{.5em}
\begin{equation}\label{eq:en}
    E_N = \frac{E_0e^{j\theta}\left(1-r_2r_Ee^{j(\phi_E+\phi_2)}\right)}{2-r_2e^{j\phi_2}(r_Ee^{j\phi_E}+r_Ne^{j\phi_N})} \;.
\end{equation}
\\Substituting \eqref{eq:a1} for $E_N$ in \eqref{eq:a2} and solving for $E_E$,
\begin{equation}\label{eq:ee}
    E_E = \frac{E_0e^{j\theta}\left(1-r_2r_Ne^{j(\phi_N+\phi_2)}\right)}{2-r_2e^{j\phi_2}(r_Ee^{j\phi_E}+r_Ne^{j\phi_N})} \;.
\end{equation}
\section{Derivation of GMI transmission (\ref{eq:et})}\label{app:b}
The field transmitted at the output port of the Grover coin, i.e. the bottom port of BS1, is a coherent superposition of the field returning from the first mirror M1, and the one returning from the input port of BS2.
\begin{equation}
    E_t = \frac{1}{2}\left[E_0r_1e^{j\phi_1}+(-r_N)e^{j\phi_N}E_N+(-r_E)e^{j\phi_E}\right]
\end{equation}
Substituting \eqref{eq:en} for $E_N$ and \eqref{eq:ee} for $E_E$, this becomes (after normalizing to $E_0$)
\begin{equation}
\begin{split}
     \frac{E_t}{E_0}= \frac{e^{j2\theta}}{2}\left[\frac{(r_Ne^{j\phi_N}+r_Ee^{j\phi_E}-2r_Nr_Er_2e^{j\Phi})}{r_2e^{j\phi_2}(r_Ne^{j\phi_N}+r_Ee^{j\phi_E})-2}\right]\\
     + \frac{r_1e^{j\phi_1}}{2}
    \end{split}
\end{equation}
where $\Phi=\phi_N+\phi_E+\phi_2$. The field reflected from the GMI is the same but with a minus sign in front of the final term due to the orientation of the dielectric beam splitter, i.e.
\begin{equation}
\begin{split}
     \frac{E_r}{E_0}= \frac{e^{j2\theta}}{2}\left[\frac{(r_Ne^{j\phi_N}+r_Ee^{j\phi_E}-2r_Nr_Er_2e^{j\Phi})}{r_2e^{j\phi_2}(r_Ne^{j\phi_N}+r_Ee^{j\phi_E})-2}\right]\\
     - \frac{r_1e^{j\phi_1}}{2} \;.
    \end{split}
\end{equation}
\section{Derivation of GMI phase function (\ref{eq:gamma})}\label{app:c}
The argument of \eqref{eq:gammadef} is defined as $\gamma(\phi_N,\phi_E) = \text{atan2}\left(\Im\{e^{j\gamma}\},\;\Re\{e^{j\gamma}\}\right)$, where $\Im\{\cdot\}$ and $\Re\{\cdot\}$ denote the imaginary and real parts of $\{\cdot\}$, respectively. The complex ratio in $e^{j\gamma}$ can be rewritten as $nd^*/dd^*$, where $n$ is the numerator and $d$ the denominator. This gives
\begin{equation}
\frac{(e^{j\phi_N}+e^{j\phi_E}-2e^{j(\phi_N+\phi_E)})(e^{-j\phi_N}+e^{-j\phi_E}-2)}{|e^{j\phi_N}+e^{j\phi_E}-2|^2}\;.    
\end{equation}
The argument then becomes $\text{atan2}\left(\Im\{nd^*\},\;\Re\{nd^*\}\right)$, since the common real denominator $dd^*$ cancels out. Using Euler's identity $2\cos x= (e^{jx}+e^{-jx})$, the numerator simplifies to
\begin{equation}
    2+2\cos{(\phi_N-\phi_E)}+4(e^{j(\phi_N+\phi_E)}-e^{j\phi_N}-e^{j\phi_E})\;.
\end{equation}
Using Euler's identity again, $e^{jx}=\cos x+j\sin x$, the nonlinear phase is
\begin{widetext}
\begin{equation}
    \gamma(\phi_N,\phi_E)=\text{atan2}
    \left(\sin\Theta-\sin\phi_N-\sin\phi_E,\;\cos\Theta-\cos\phi_N-\cos\phi_E+\frac{1}{2}(1+\cos \phi)\right)
\end{equation}
\end{widetext}
where $\Theta=\phi_N+\phi_E$ and $\phi=\phi_N-\phi_E$. 
\section{GMI transmission maxima}\label{app:tmax}
Since the transmitted intensity is proportional to $\cos^2 (\gamma/2)$, there will be transmission maxima whenever the numerator in \eqref{eq:gamma} takes on a value of 0 mod 2$\pi$. Using the identity $\sin(x+y)=\sin x\cos y+\cos x \sin y$, the numerator can be rewritten as
\begin{equation}
\begin{split}
    &\sin\phi_N\cos\phi_E+\cos\phi_N\sin\phi_E-\sin\phi_N-\sin\phi_E\\
    & = \sin\phi_N(\cos\phi_E-1)+\sin\phi_E(\cos\phi_N-1)\;.
    \end{split}
\end{equation}
Applying the identity $\cos(2x)-1=-2\sin^2x$, we have
\begin{equation}
    -2\sin\phi_N\sin^2\left(\frac{\phi_E}{2}\right)-2\sin\phi_E\sin^2\left(\frac{\phi_N}{2}\right)
\end{equation}
The identity $\sin (2x) = 2\sin x\cos x$ yields
\begin{widetext}
\begin{equation}
\begin{split}
    &-4\sin\left(\frac{\phi_N}{2}\right)\cos\left(\frac{\phi_N}{2}\right)\sin^2\left(\frac{\phi_E}{2}\right)
    -4\sin\left(\frac{\phi_E}{2}\right)\cos\left(\frac{\phi_E}{2}\right)\sin^2\left(\frac{\phi_N}{2}\right)\\
     &= -4\sin\left(\frac{\phi_N}{2}\right)\sin\left(\frac{\phi_E}{2}\right)
    \left[\sin\left(\frac{\phi_E}{2}\right)\cos\left(\frac{\phi_N}{2}\right)+\sin\left(\frac{\phi_N}{2}\right)\cos\left(\frac{\phi_E}{2}\right)\right]\;.
    \end{split}
\end{equation}
\end{widetext}
Implementing the first identity again yields the condition for transmission extrema,
\begin{equation}
    -4\sin\left(\frac{\phi_N}{2}\right)\sin\left(\frac{\phi_E}{2}\right)\sin\left(\frac{\phi_N+\phi_E}{2}\right)=0\;.
\end{equation}
The trivial solutions $\phi_N=0\,\text{mod} \,2\pi$ and $\phi_E=0\,\text{mod}\,2\pi$ are spurious zero crossings in $\gamma$; they produce $\gamma=0$ but physically constitute destructive interference at the output port with the amplitude returning to BS1 from M1. This is seen in \ref{eq:et} where, setting $\phi_E=0$ for a finite $\phi_N$, the transmitted field is
\begin{equation}
    \tilde{E}_t=\frac{1}{2}\left[\frac{1-e^{j\phi_N}}{e^{j\phi_N}-1}+1\right]=0\;.
\end{equation}
The condition $\phi_E=-\phi_N$ produces the transmission maxima observed in Fig. \ref{fig:gamma}, where the two cavities contribute to the resonance symmetrically. However, it is important to note that this is in the case of balanced beam splitters and round-trip phases of $\phi_1$, $\phi_2$, and $\theta$ all set to values of $0\,\text{mod}\,2\pi$. A shift in any of the three produces a global phase shift of the nonlinear phase level sets, and imbalanced beam splitters result in reduced visibility.

\bibliography{bibliography}

\end{document}